\documentclass[english,aps,twocolumn,pre,epsfig,superscriptaddress,showpacs,amsmath,amsfonts,amssymb,floatfix]{revtex4-1}
\usepackage{amsmath,amsfonts,amssymb,color}
\usepackage{graphicx}
\usepackage{dcolumn}
\usepackage{bm}
\usepackage[latin9]{inputenc}
\usepackage{mathtools}
\usepackage{amssymb}
\usepackage{esint}
\usepackage{babel}

\usepackage{bbm}

\begin{document}
\title{Merits and Qualms of Work Fluctuations in Classical Fluctuation Theorems}
\author{Jiawen Deng}
\affiliation{NUS Graduate School for Integrative Science and Engineering, Singapore 117597}
\author{Alvis Mazon Tan}
\affiliation{Department of Physics, National University of Singapore, Singapore 117546}
\author{Peter H\"{a}nggi}
\affiliation{Department of Physics, National University of Singapore, Singapore 117546}
\affiliation{Institute of Physics, University of Augsburg, Universit\"{a}tsstra$\beta${\it e 1, D-86135} Augsburg, Germany}
\author{Jiangbin Gong} \email{phygj@nus.edu.sg}
\affiliation{Department of Physics, National University of Singapore, Singapore 117546}
\affiliation{NUS Graduate School for Integrative Science and Engineering, Singapore 117597}
\date{\today}




\begin{abstract}
Work is one of the most basic notion in statistical mechanics, with work fluctuation theorems being one central topic in nanoscale thermodynamics.  With Hamiltonian chaos commonly thought to provide a foundation for classical statistical mechanics, here we present general salient results regarding how (classical)  Hamiltonian chaos generically impacts on nonequilibrium work fluctuations. For isolated chaotic systems prepared with a microcanonical distribution, work fluctuations are minimized and vanish altogether in adiabatic work protocols. For isolated chaotic systems prepared at an initial canonical distribution at inverse temperature $\beta$,  work fluctuations depicted by the variance of $e^{-\beta W}$ are also minimized by adiabatic work protocols. This general result indicates that, if the variance of $e^{-\beta W}$ diverges for an adiabatic work protocol,
it diverges for all nonadiabatic work protocols sharing the same initial and final Hamiltonians.
Such divergence  is hence not  an isolated event and thus greatly impacts on the efficiency of using the Jarzynski's equality to simulate free energy differences. Theoretical results are illustrated in a Sinai model.
Our general insights shall boost studies in
nanoscale thermodynamics and are of fundamental importance in designing useful work protocols.
\end{abstract}


\maketitle

\section{Introduction}
Within the realm of classical mechanics of closed many-body systems, the ergodic hypothesis of equilibrium statistical mechanics  (SM) and related issues such as  thermalization and equilibration, as pioneered by Gibbs, Maxwell, Boltzmann, Caratheodory and others \cite{sklar1993},
is closely connected with an underlying Hamiltonian chaotic dynamics.
Indeed, chaos-induced ergodicity renders equilibrium SM concepts applicable to chaotic systems with few degrees of freedom \cite{quartic1,quartic2,quartic3}.  Even more fundamental, chaos is crucial towards understanding nonequilibrium SM (for systems large and small) \cite{book1}, because it offers a potential answer to the emergence of a time arrow \cite{Tian} and is essential also in understanding diffusion, conduction, thermalization processes \cite{Casati2,Casati3,lepri2016,Izrailev}, etc.

Given the fundamental connection between chaos and SM,  here we aim to reveal a number of generic features of classical work fluctuations in nonequilibrium processes in chaotic systems, regardless of the number of degrees of freedom.  We note that work is one of the most basic notions in SM and their fluctuation aspects have been the central topic in seminal fluctuation theorems \cite{bochkovkuzolev1977,JarzynskiPRL,Crooks_Relation,ColloquiumRevModPhys.83.771}.  General understandings of work fluctuations in chaotic systems can further boost studies of nanoscale thermodynamics. They can also guide future energy device designs such as heat engines operating at the nanoscale \cite{enginePRL,DengPRE,Campo,Scienceengine,review1,review2,review3},  where work fluctuations should be suppressed to achieve a more uniform work output and a higher heat-to-work effciency \cite{GaoyangPRE15}.

We shall reveal that ergodicity arising from Hamiltonian chaos in isolated systems has  far-reaching statistical implications.
For initial states prepared as a microcanonical distribution, work fluctuations vanish {\it identically} in the (mechanical) adiabatic limit.  For initial states prepared with a canonical distribution at the inverse Boltzmann temperature $\beta$ \cite{hanggi_PRE,RSTA}, the variance in $e^{-\beta W}$ as an exponential form of work $W$, namely,
 ${\rm Var}(e^{-\beta W})\equiv \langle e^{-2\beta W}\rangle - \langle e^{-\beta W}\rangle^2$,
is minimized by adiabatic work protocols.  This second result indicates that if ${\rm Var}(e^{-\beta W})$ diverges
for an adiabatic protocol, then it diverges for any nonadiabatic work protocol sharing the same initial and final
Hamiltonians. As such, divergence in ${\rm Var}(e^{-\beta W})$ may occur systematically.
 In some previous studies the error in free-energy simulations based on Jarzynski's equality was analyzed based on a finite
${\rm Var}(e^{-\beta W})$ \cite{Jarzynski2,Dellago2}. A diverging ${\rm Var}(e^{-\beta W})$ immediately suggests a challenge.
As shown below,
 the systematic divergence of ${\rm Var}(e^{-\beta W})$ greatly impacts on the convergence of simulated averages of $e^{-\beta W}$ (over a statistical sample of finite size) towards its theoretical mean value $e^{-\beta \Delta F}$ via Jarzynski's equality \cite{JarzynskiPRL}
(where $\Delta F$ represents a free energy difference). This implication for the efficiency of Jarzynski's equality shall stimulate immediate theoretical and experimental interests.



\section{Work fluctuations: Isolated systems prepared at microcanonical equilibrium}
In our considerations below, a Hamiltonian system parameterized by an external control parameter $\lambda$
is assumed to be completely chaotic to induce ergodicity on the energy surface (here we only need the ergodicity aspect of chaos).
Let $\Omega(E,\lambda)$ be the properly normalized (via division by symmetry factors)  and dimensionless (via division by the appropriate power of Planck's constant) phase space volume enclosed by the energy surface at energy $E$, i.e.
\begin{equation}
\Omega(E;\lambda)=\int_{\Gamma}\Theta\big(E-H({\bf p},{\bf q};\lambda)\big){\rm d}{\bf p}{\rm d}{\bf q},
\label{omega}
\end{equation}
where $\Theta$ is the unit step function, $({\bf p},{\bf q})$ are a collection of
phase space variables, and $H({\bf p},{\bf q};\lambda)$ is the chaotic Hamiltonian. In this form it equals the integral of the density of states (DoS), i.e.,
$\Omega(E;\lambda)= \int_{E_{\text{min}}}^E \omega(E;\lambda) dE$ with the  DoS given by $\omega(E;\lambda)= {\partial \Omega(E;\lambda)}/{\partial E}$.
For a work protocol implemented by time-varying $\lambda(t)$, the energy change associated with
$\lambda\rightarrow \lambda +{\rm d}\lambda$ is given by
\begin{equation}
{\rm d}H=\frac{\partial H}{\partial t} {\rm d} t = \frac{\partial H}{\partial \lambda} {\rm d}\lambda.
\end{equation}
If the rate of change in $\lambda$ is mechanically slow (adiabatic), then the time
it takes to realize the change of ${\rm d}\lambda$ is very long, during which the trajectory
manifests ergodicity; i.e.,   it visits the entire energy surface with equal probability. We then obtain
\begin{equation}
{\rm d}H=\left\langle\frac{\partial H}{\partial \lambda}\right\rangle_{E} {\rm d} \lambda\;,
\label{ereq}
\end{equation}
where $\langle\cdot\rangle_E$ represents microcanonical average over an energy surface at energy $E$.
 As seen from Eq.~(\ref{ereq}), the energy change becomes {\it independent} of the initial condition on the starting energy surface.  Then, for a work protocol starting out from
$\lambda(t=0)=\lambda_0$ with initial conditions $({\bf p}_{0},{\bf q}_{0})$ and ending
 with $\lambda(t=\tau)=\lambda_{\tau}$ with final states $({\bf p}_{\tau},{\bf q}_{\tau})$,
 the final energy for mechanically adiabatic driving as a function of initial conditions $({\bf p}_{0},{\bf q_0};\lambda_0)$, i.e., $E_{\tau}({\bf p}_{0},{\bf q}_{0};\lambda_0)$, becomes  {\it independent} of  the set of all possible initial conditions $({\bf p}_{0},{\bf q}_{0})$  starting from the energy surface $H({\bf p}_{0},{\bf q}_{0};\lambda_0)= E_0$.
 The inclusive work \cite{JarzynskiWork,ColloquiumRevModPhys.83.771}
 \begin{equation}
 W=H({\bf p}_{\tau},{\bf q}_{\tau};\lambda_\tau)-H({\bf p}_{0},{\bf q}_{0};\lambda_0)= E_{\tau}-E_0 \;.
 \label{work}
\end{equation}
thus becomes  {\it fixed} at the value $W_{\text{ad}}\equiv E_{\tau} - E_{0}$.

Consequently,  the work fluctuation vanishes identically if we start with a microcanonical ensemble preparation.  Indeed, all the possible final states end up on a new energy surface corresponding to a new microcanonical ensemble. By use of a more rigorous analysis and in connection with Liouville's theorem,  one recovers the salient result, derived earlier in   the literature \cite{Hertz,USSR,kasuga1961_1,adiabatic_invariant_1,adiabatic_invariant_2,sasa}, namely, the phase space volume $\Omega(E_\tau;\lambda_\tau)$ equals $\Omega(E_0;\lambda_0)$ if the protocol is mechanically adiabatic. This should not be confused with the better known Liouville theorem, where phase space volume is always preserved but there is no guarantee that the final states are on the same energy surface.  Thus, $\Omega(E;\lambda)$ in Eq.~(\ref{omega}) is a thermodynamic adiabatic invariant \cite{hanggi_PRE,RSTA}. The remarkable result that work fluctuations in chaotic systems vanish identically in the adiabatic limit is seen to share the same physics behind the ergodic adiabatic invariant $\Omega(E; \lambda)$.

\begin{figure} 
\centering      
\includegraphics[width=0.8\linewidth]{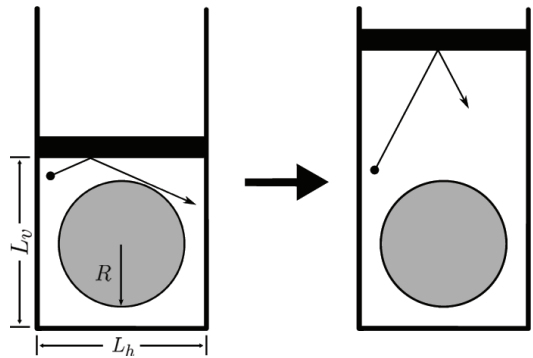}
\caption{The chaotic Sinai billiard with a moving piston (via changing $L_v$).
$\lambda$ in the main text refers to
the free area enclosed by the billiard.  For comparison, a rectangular billiard without the circular structure of radius $R$ is also considered as a non-chaotic test case. We set $L_h=40$ and $R=15$,
with $L_v=40$ at the beginning of all work protocols.}
\label{fig:sinai}
\end{figure}

Note in passing that the above finding supplements earlier theorems for work fluctuations for a microcanonical preparation \cite{JouleExperiment,hanggimicro1,campisi,hanggimicro2}. Those early theorems showed that the ratio of  two  work distributions for forward and backward protocols, {\it independent of the time-variation of $\lambda(t)$}, equals the ratio of two DoS, namely  $ \omega(E_0 + W)/\omega(E_0)$.  Now given that  $\Omega(E;\lambda)$ is an adiabatic invariant, i.e., $\Omega(E_{\tau};\lambda_{\tau}) = \Omega(E_0;\lambda_0)$, this ratio reduces to that of corresponding two thermodynamic Gibbs temperatures $k_B T_G= \Omega (\cdot)/\omega(\cdot)$ \cite{hanggi_PRE,RSTA} because  $\omega(E_{\tau}; \lambda_{\tau})/\omega(E_{0};\lambda_0) = T_{G}(E_0;\lambda_0)/ T_{G}(E_{\tau};\lambda_{\tau})$.

\begin{figure} 
\centering      
\includegraphics[width=0.7\linewidth]{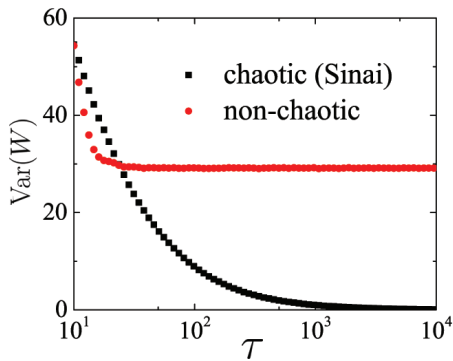}
\caption{(color online) Variance of work fluctuations, denoted Var$(W)$,
 vs the duration $\tau$ of the work protocol $\lambda_0\rightarrow \lambda_\tau=1.2\lambda_0$, obtained from $2.4\times 10^5$ individual work realizations. Regimes with very large $\tau$ refer to adiabatic situations.
For the Sinai billiard case, Var$(W)$ approaches $0$ as $\tau$ goes to infinity;
 For the non-chaotic case (square billiard) with the same $\lambda_\tau/\lambda_0$,
 Var$(W)$ stays well above zero even for large $\tau$.
Here and elsewhere of the main text, all plotted quantities are scaled and hence dimensionless, with $E_0=50$, and $m=1$. }
\label{fig:micro-canonical}
\end{figure}

As a computational example throughout this work, consider  a modified 2D Sinai billiard \cite{Sinai} depicted in Fig. \ref{fig:sinai}.  The billiard has a moving wall as an analog of a piston.  The ergodic adiabatic invariant $\Omega(E;\lambda)$ is found to be $\Omega(E;\lambda)=2\pi m E\lambda$ where $\lambda$ in this case represents the area of the billiard, $m$ is the mass of the point particle inside the billiard.  In the adiabatic limit, all states starting from a microcanonical ensemble at $E=E_0$ have the same final
energy $E_{\tau}=(\lambda_{0}/\lambda_{\tau})E_0$, resulting in vanishing work fluctuations.
By contrast, in an analogous setup but without the circular pillar (that is, just a rectangular billiard), the system is integrable. Ergodicity then no longer holds. Using that
the piston moving direction decouples from the transverse direction, one can theoretically obtain that the variance in $W$ is always above zero, even when the piston moves adiabatically.
Our numerical simulations depicted in Fig.~\ref{fig:micro-canonical}
confirm these insights. In particular, (throughout) we consider a work protocol $\lambda(t)=\lambda_0+(\lambda_\tau-\lambda_0)[1-\cos(\pi t/\tau)]/2$ of duration $\tau$ \cite{note2}.
It is seen that the variance for $W$
decreases to $0$ for the chaotic Sinai billiard; in contrast it approaches  a nonzero value (in agreement with theory)  for the integrable billiard.

\section{Work fluctuations: Isolated systems prepared at canonical equilibrium} Jarzynski's equality $\langle e^{-\beta W}\rangle=e^{-\beta\Delta F}$ depicts
nonequilibrium work fluctuations in systems initially prepared as Gibbs states
at inverse Boltzmann temperature $\beta$. Here, $\langle\cdot \rangle$ now denotes a statistical average over the initial canonical distribution.  Going beyond this, it is of interest to ask
how individual realizations of $e^{-\beta W}$ obtained in experiments or simulations
deviate from their mean value $e^{-\beta\Delta F}$.

Once the initial and final Hamiltonians are specified by the
parameters $\lambda_0$ and $\lambda_\tau$, then $\Delta F$ as well as $\langle e^{-\beta W}\rangle$ become fixed for all possible work protocols $\lambda(t)$ that start with $\lambda_0$ and end at $\lambda_\tau$,   yielding  ${\rm Var}(e^{-\beta W})=  \langle e^{-2\beta W}\rangle - e^{-2\beta \Delta F}$. We next seek to minimize ${\rm Var}(e^{-\beta W})$ among all possible such work protocols. Note that the mapping between the adiabatic invariant $\Omega(E; \lambda)$ and the
energy $E$ is injective for any given $\lambda$.  We thus write $E=E(\Omega;\lambda)$ with the adiabatic invariant $\Omega$ acting as the ``indicator" for the corresponding energy surface. For an arbitrary (nonadiabatic in general) work protocol,
these relations between energy and phase space volume can still be used, e.g., $\bar{E}_0 = E(\bar{\Omega}_0;\lambda_0)$, $\bar{E}_{\tau}=E(\bar{\Omega}_{\tau};\lambda_{\tau})$, with $\bar{\Omega}_{0} = \Omega(\bar{E}_0;\lambda_0)$,   $\bar{\Omega}_{\tau} = \Omega(\bar{E}_\tau;\lambda_\tau)$, and the bar referring to quantities associated with a general work protocol.  We next make use of a useful relation, reading for  any phase space integral of an arbitrary function $f\big(H({\bf p},{\bf q};\lambda)\big)$  with the energy $E(\Omega; \lambda)$ not being bounded from above (see Appendix A):

\begin{equation}
\int_{\Gamma}f\big(H({\bf p},{\bf q};\lambda)\big){\rm d}{\bf p}{\rm d}{\bf q}=\int_{0}^{\infty}f\big(E(\Omega;\lambda)\big){\rm d}\Omega\;.
\end{equation}
Here, the lower boundary  $\Omega=0$ corresponds to the vanishing phase space volume, cf Eq.~(\ref{omega}), at  bounded lowest energy $H({\bf p},{\bf q};\lambda)$ at fixed control parameter $\lambda$.
Particularly, work statistics over different initial states
$({\bf p}_0,{\bf q}_0)$ and different final states $({\bf p}_\tau,{\bf q}_\tau)$ can be now transferred to statistics over phase space volumes.  For a system initially at inverse temperature $\beta$ and subject to an arbitrary  time-varying work protocol $\lambda(t)$ with $\lambda_{0}\rightarrow \lambda_{\tau}$,  the quantity $\langle e^{-2\beta W}\rangle$ can  be recast with Eq.~(\ref{work}) as
\begin{eqnarray}
\label{aveexpW}
\langle e^{-2\beta W}\rangle &=&\int_{0}^{\infty}{\rm d}\bar{\Omega}_{\tau}\int_{0}^{\infty}{\rm d}\bar{\Omega}_0\; e^{-2\beta[E(\bar{\Omega}_{\tau};\lambda_{\tau})-E(\bar{\Omega}_0;\lambda_{0})]}\nonumber \\
&&\times\ P(\bar{\Omega}_{\tau}|\bar{\Omega}_0)\rho_{0}(\bar{\Omega}_0)\;,
\end{eqnarray}
where $\bar{\Omega}_0$ represents the phase space volume enclosed by an initial energy value $\bar{E}_0=E(\bar{\Omega}_0;\lambda_0)$, the factor $\rho_{0}(\bar{\Omega}_0;\lambda_{0})=e^{-\beta E(\bar{\Omega}_0;\lambda_{0})}/Z(\beta;\lambda_{0})$
arises from the initial canonical Gibbs distribution, $Z(\beta;\lambda_{0})$ is the partition function associated with Hamiltonian $H({\bf p},{\bf q};\lambda_0)$, $\bar{\Omega}_{\tau}$ represents the phase space volume enclosed by the final energy value $\bar{E}_{\tau}=E(\bar{\Omega}_{\tau};\lambda_\tau)$.
Notably, the part   $P(\bar{\Omega}_{\tau}\vert\bar{\Omega}_0)$ in Eq.~(\ref{aveexpW}) is the conditional probability
 describing the transition for the system to start from $\bar{\Omega}_0$ and to end up with $\bar{\Omega}_{\tau}$. As shown in Appendix B, this conditional probability is bi-stochastic, obeying

\begin{equation}
\int_{0}^{\infty}P(\bar{\Omega}_{\tau}|\bar{\Omega}_0){\rm d}\bar{\Omega}_{\tau}= 1 =\int_{0}^{\infty}P(\bar{\Omega}_0|\bar{\Omega}_{\tau}){\rm d}\bar{\Omega}_{\tau}.\label{eq:bi-stochastic}
\end{equation}

Based on the above relation, we can prove (See Appendix C) that
$\langle e^{-2\beta W}\rangle$ or equivalently ${\rm Var}(e^{-\beta W})$ is minimized if $P(\bar{\Omega}_{\tau}|\bar{\Omega}_0)=\delta(\bar{\Omega}_{\tau}-\bar{\Omega}_0)$. Put differently, minimization of $\langle e^{-2\beta W}\rangle$ is realized if the phase space volume enclosed by an arbitrary initial energy surface at $\bar{E}_0=H({\bf p}_0,{\bf q}_0;\lambda_0)$
 {\it equals}
the phase space volume enclosed by the energy surface at the final energy $\bar{E}_\tau=H({\bf p}_\tau,{\bf q}_\tau;\lambda_\tau)$.
This salient condition can be just achieved using an {\it adiabatic} work protocol, as detailed above.
In short, for all work protocols $\lambda(t)$ with  $\lambda_{0}\rightarrow \lambda_{\tau}$ we find the lower bound
$\langle e^{-2\beta W_{\text{ad}}}\rangle\le\langle e^{-2\beta W}\rangle$, which indicates
\begin{equation}
{\rm Var}(e^{-\beta W_{\text{ad}}}) \le {\rm Var}(e^{-\beta W})  \;.
\label{eq:minimal_second_moment}
\end{equation}

The above result extends a previous principle of minimal {\it exponential} work fluctuations ~\cite{Gaoyang_Minimal_PhysRevE.92.022130}, established for strictly integrable   systems and in effect only applicable to one-dimensional Hamiltonians (as a consequence of those limiting prerequisites stated in Ref.~\cite{Gaoyang_Minimal_PhysRevE.92.022130}), to chaotic systems with an arbitrary number of degrees of freedom. It is the feature of underlying full chaos and hence ergodicity that makes this extension possible. Specifically, phase space volume $\Omega$ serves as an index for energy surfaces and
ergodicity makes $\Omega$ an adiabatic invariant even in multi-dimensional systems.
This leads to a preserved energy surface index and adiabatic work protocols then stand out from all other work protocols.

\begin{figure} 
\centering      
\includegraphics[width=0.7\linewidth]{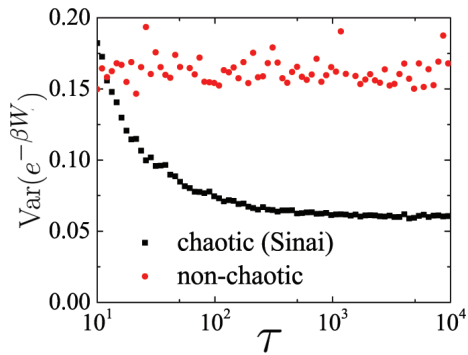}
\caption{(color online) Work fluctuations depicted by the variance of $e^{-\beta W}$ vs the duration $\tau$ of the work protocol $\lambda_0\rightarrow \lambda_\tau=1.2\lambda_0$, with $2.4\times 10^5$ individual work realizations  and $\beta=0.01$.
In the chaotic case,  the variance of $e^{-\beta W}$ reaches its
 minimal values as $\tau$ becomes very large (adiabatic regime), in agreement with theory. In the non-chaotic case,
the variance of $e^{-\beta W}$  oscillates violently {\it vs.} $\tau$ \cite{footnote}. Remaining parameters are as for Fig. \ref{fig:micro-canonical}.}
\label{fig:canonical}
\end{figure}

Let us next test these predictions via computational results for the Sinai billiard, as compared with
the integrable billiard case. As depicted in Fig.~\ref{fig:canonical}, with the increase
of the duration $\tau$ of a work protocol $\lambda_0\rightarrow \lambda_{\tau}$,  ${\rm Var}(e^{-\beta W})$
in the chaotic Sinai billiard case decreases monotonically and minimization of
${\rm Var}(e^{-\beta W})$ is achieved as $\tau \rightarrow \infty$, reaching the adiabatic regime.
By contrast, based on results from $2.4\times 10^5$ non-chaotic trajectories (integrable billiard),
we detect  no obvious trend of a systematic decrease in ${\rm Var}(e^{-\beta W})$,  no matter how large $\tau$ becomes \cite{footnote}.
 It is also of interest to construct a specific example where ${\rm Var}(e^{-\beta W})$ actually increases
as $\tau$ increases in a non-ergodic system.  This is not possible for the integrable billiard case here because the piston degree of freedom can be completely decoupled from its transverse motion.  Instead, in Appendix D we offer one such specific example using a nonlinear two-dimensional oscillator system with a mixed classical phase space.

 \section{Divergence of ${\rm Var}(e^{-\beta W})$} We now reach the key section of this study. Exploiting the adiabatic invariant $\Omega(E; \lambda)$, one may also investigate if an upper bound of ${\rm Var}(e^{-\beta W_{\text{ad}}})$ exists. Inverting the relation $\Omega(E_0;\lambda_0) = \Omega(E_{\tau}; \lambda_{\tau})$, the final energy $E_{\tau}$ after an adiabatic work protocol
can be expressed solely as a  function  $E_0, \lambda_0, \lambda_{\tau}$, denoted as $E_{\tau} = \mathcal {Y}(E_0; \lambda_0, \lambda_{\tau})$.  One then has
\begin{eqnarray}
\label{aveexpW2}
\langle e^{-2\beta W_{\text{ad}}}\rangle &=& \int_{0}^{\infty}e^{- \beta[2\mathcal{Y}(E_0;\lambda_0, \lambda_{\tau})-E_0]}\frac{\omega(E_0;\lambda_0)}{Z(\beta;\lambda_0)}{\rm d}E_{0}.\ \
%
%
\end{eqnarray}
As increasing values
$E_0$ are sampled from the initial Gibbs state, there is no reason to expect that the integrand $e^{-\beta[2\mathcal{Y}(E_0,s_0, \lambda_{\tau})-E_0]}$ should be always bounded from above.
This being the case, $\langle e^{-2\beta W_{\text{ad}}}\rangle$ and hence
the associated ${\rm Var}(e^{-\beta W_{\text{ad}}})$ can diverge (no upper bound).  This general possibility of divergence
in ${\rm Var}(e^{-\beta W_{\text{ad}}})$ goes beyond an earlier observation for the adiabatic expansion of a one-dimensional ideal gas \cite{Dellago1} to chaotic systems with an arbitrary number of degrees of freedom. More significantly,
according to Eq.~(\ref{eq:minimal_second_moment}), ${\rm Var}(e^{-\beta W})$ then diverges for all nonadiabatic work protocols $\lambda_0\rightarrow\lambda_\tau$ of arbitrary duration $\tau$ (that is, divergence may occur systematically!)

For the Sinai billiard, $\Omega(E;\lambda)=2\pi mE\lambda$, $E_\tau=\mathcal{Y}(
E_0, \lambda_0,\lambda_\tau)=\frac{\lambda_0}{\lambda_\tau}E_0$, we obtain
\begin{eqnarray}
\label{eq:divergent_second}
\langle e^{-2\beta W_{\text{ad}}}\rangle &=& \frac{2\pi m \lambda_0}{Z(\beta;\lambda_{0})}\int_{0}^{\infty}e^{-\beta(\frac{2\lambda_0}{\lambda_{\tau}}-1) E_0}  {\rm d}E_{0}\nonumber \\
 & = & \frac{\lambda_{\tau}}{2\lambda_0-\lambda_{\tau}}\ \ \  (\text{if}\ \lambda_{\tau} <2\lambda_0).
\end{eqnarray}
This main result Eq.~(\ref{eq:divergent_second}) shows that if $\lambda_\tau \geq 2 \lambda_0$, then the positive-definite
 quantity $\langle e^{-2\beta W_{\text{ad}}}\rangle$ diverges, resulting in the divergence
 of ${\rm Var}(e^{-\beta W})$ for all nonadiabatic protocols with $\lambda_\tau \geq 2 \lambda_0$!

The efficiency of using Jarzynski's equality to simulate $e^{-\beta \Delta F}$ from averaging
over $n$ realizations of $e^{-\beta W}$ is of great practical interest \cite{Jarzynski2,Dellago2,Kim,Hartmann}.
For a diverging ${\rm Var}(e^{-\beta W})$,  the familiar central limit theorem (CLT) can no longer predict how the simulation
error scales with the sample size $n$. As such, even an error analysis becomes a challenge.
Here we use a generalization of CLT \cite{durrett,kolmogorov}
to arrive at a specific prediction,  which is the first of such kind.

Let $s \equiv  {\lambda_\tau}/{\lambda_{0}}$. For those cases with a diverging ${\rm Var}(e^{-\beta W_{\text{ad}}})$
we have $s \ge 2$. Consider the $95\%$ confidence interval (CI) width of the average $\sum_{i=1}^{n}e^{-\beta W_{i}}/n$ over a statistical sample of size $n$.
The CI width (roughly called error below) is about the spread
of the distribution of $\sum_{i=1}^{n}e^{-\beta W_{i}}/n$, obtained by repeating simulations or experiments based on $n$ work realizations.
The smaller the CI width is, the more accurate is the simulated $e^{-\beta \Delta F}$. With details presented in Appendix E, we discover that the scaling law of the CI width with $n$ is given by $\sim n^{-\frac{1}{s}}$ for $s\ge2$, which is markedly different from the familiar CLT scaling $1/\sqrt{n}$ (unless $s=2$).
For example, if $s=3$, then the error of $\sum_{i=1}^{n}e^{-\beta W_{i}}/n$
scales as $\sim n^{-\frac{1}{3}}$.   That is, if we hope to increase the accuracy by one decimal place, the case of $s=3$ with diverging ${\rm Var}(e^{-\beta W})$ would already need to increase the sample size by 1,000 times. By contrast, in the normal situation one only needs to increase the sample size by 100 times.
 Further, if $\lambda_\tau\gg \lambda_0$, then $1/s$ approaches zero, yielding an extremely slow error scaling. In these cases Jarzynski's equality would
 be {\it impractical} in simulating the free energy difference for both adiabatic and nonadiabatic work protocols.

\begin{figure} 
\centering      
\includegraphics[width=0.7\linewidth]{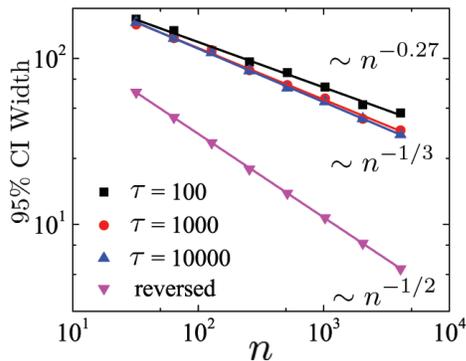}
\caption{(color online) The 95\% confidence interval width of the simulated average $\sum_{i=1}^{n}e^{-\beta W_{i}}/n$ over $2.4\times 10^4$ statistical samples, as a function of the sample size $n$, with $\beta=0.01$. For the bottom curve, $s=\lambda_\tau/\lambda_0=1/3$ and so ${\rm Var}(e^{-\beta W})$ being finite, the error scaling is in agreement with CLT.
For the upper three curves, $s=3$ and hence ${\rm Var}(e^{-\beta W})$  diverges, with the error scaling given by
$\sim n^{-\frac{1}{3}}$ in the adiabatic case ($\tau=10^4$) or even slower scalings in the other two cases. }
\label{fig:stable_law}
\end{figure}

Computational results in Fig.~\ref{fig:stable_law} confirm our theoretical analysis. There we consider cases with $s=3$, for both adiabatic and nonadiabatic situations.  The results are also compared with a reversed adiabatic work protocol, namely, $s=1/3$, for which there is no divergence
in ${\rm Var}(e^{-\beta W_{\text{ad}}})$.  Several observations can be made. First, for the adiabatic work protocol ($\tau=10^4$) with $s=1/3$, the statistical error
represented by the bottom line indeed scales with $n^{-\frac{1}{2}}$, consistent with the common CLT.
Second, for the adiabatic work protocol ($\tau=10^4$) with $s=3$,  the width of 95\% CI of $\sum_{i=1}^{n}e^{-\beta W_{i}}/n$ scales as $n^{-\frac{1}{3}}$, in perfect agreement with the prediction above. This also indicates that
they have totally different efficiency in simulating  $e^{-\beta \Delta F}$.
Third, the upper two curves (red dots and black squares) in Fig.~\ref{fig:stable_law} depict the scaling of the error for nonadiabatic situations with $s=3$ and hence diverging ${\rm Var}(e^{-\beta W})$.  The error scaling for the highly nonadiabatic case ($\tau=100$) is even slower than $n^{-\frac{1}{3}}$. Thus, in the presence of a diverging ${\rm Var}(e^{-\beta W})$, the adiabatic work protocol is still much more beneficial over using a highly nonadiabatic work protocol because the former possesses advantageous error scaling.


 \section{Concluding Remarks} Aspects of work fluctuations continue to be of fundamental interest \cite{Hanggi_work_PRE,biao,gaoyangPRE,Dario}.
Deep insights into work fluctuations are also relevant to designs of energy devices.
Our general results would not be possible were the system under consideration not chaotic (systems with only one degree of freedom can be exceptions \cite{hanggi_PRE} because they are typically ergodic).
This study shows that, when suppression of work fluctuations is of interest, the underlying chaotic dynamics affords general predictions and hence may be of expedient value in designing work protocols.
Though derived in the classical domain, our results may invigorate the community
to research new aspects of work fluctuations and chaos in the quantum domain.    In particular, it is now urgent
to revisit Jarzynski's equality in both classical and quantum domains to arrive at some general system characteristics and work protocol guidelines to best avoid the divergence of the second moment of exponential work. We also call for next-generation strategies to simulate free energy differences via nonequilibrium work protocols.
\\

\begin{acknowledgements}  We would like to thank Gaoyang Xiao for helpful discussions.
J.D. thanks Prof. Sun Rongfeng from NUS Department of Mathematics for helpful discussions on the generalized CLT.
J.G. and P.H. are partly supported by Singapore MOE Academic Research Fund Tier-2 project (Project No. MOE2014-T2-2-
119, and J.G. with WBS No. R-144-000-350-112). The computational work was made possible by
National Supercomputing Centre, Singapore (https://www.nscc.sg).
\end{acknowledgements}

\appendix    \vspace{1cm} \begin{center} {\bf Appendix} \end{center}

This Appendix contains five sections. In section A, we prove a most useful, general equality involving an  integration over the whole phase space.  In Section B we prove the bi-stochastic property of the conditional probability $P(\bar{\Omega}_{\tau}|\bar{\Omega}_{0})$ for Hamiltonians that are not bounded from above. Section C shows that an adiabatic work protocol (applied to ergodic systems)  minimizes $\langle e^{-2\beta W}\rangle$ and hence minimizes ${\rm Var}(e^{-\beta W})$ among all work protocols sharing the same initial and final Hamiltonian.  In Section D, we discuss the behavior of ${\rm Var}(e^{-\beta W})$ in a non-ergodic system. In the last section, we discuss in greater detail the implications of a diverging ${\rm Var}(e^{-\beta W})$ by use of  a generalized central limit theorem.

\vspace{0.5cm}

\section{Useful relation involving integration over whole phase space}

For a Hamiltonian system $H({\bf p},{\bf q};\lambda)$ with generally multidimensional phase space degrees of freedom $({\bf p},{\bf q})$ and a control parameter
$\lambda$, the corresponding phase space volume up to an energy  $E$ is defined as

\begin{equation}
\Omega(E;\lambda)=\int_{\Gamma}\Theta\big(E-H({\bf p},{\bf q};\lambda)\big){\rm d}{\bf p}{\rm d}{\bf q}\;,
\label{eq:define_Omega}
\end{equation}
where $\Theta$ is the step function.
Next, we can use $\Omega\big(H({\bf p},{\bf q};\lambda);\lambda\big)$ to determine
the energy shell in phase space which contains the set of all phase space points at given energy $E$; i.e. the surface of phase points $\{({\bf p},{\bf q})\}$ obeying $E = H({\bf p},{\bf q};\lambda)$. In addition, because $\Omega$ monotonically increases with $E$, we obtain the  bijective relation that  $E=E(\Omega;\lambda)$ as a function of $\Omega$ at fixed  $\lambda$.

\begin{widetext}
For systems with a lower energy bound $E_{\text{min}}$ and unbounded energy, we can write
\begin{eqnarray}
\int_{\Gamma}f\big(H({\bf p},{\bf q};\lambda)\big){\rm d}{\bf p}{\rm d}{\bf q} & = & \int_{E_{\text{min}}}^{\infty}{\rm d}E \int_{\Gamma}f\big(H({\bf p},
{\bf q};\lambda)\big)\delta\big(E-H({\bf p},{\bf q};\lambda)\big){\rm d}{\bf p}{\rm d}{\bf q}  \\
 & = & \int_{E_{\text{min}}}^{\infty}{\rm d}Ef(E)\int_{\Gamma}\delta\big(E-H({\bf p},{\bf q};\lambda)\big){\rm d}{\bf p}{\rm d}{\bf q}  \\
 & = & \int_{E_{\text{min}}}^{\infty}{\rm d}Ef(E)\omega(E;\lambda).\label{eq:partA_01}
\end{eqnarray}
\end{widetext}
where $\omega(E;\lambda)=\partial\Omega/{\rm \partial}E$ equals the density
of states, obeying $\omega(E;\lambda){\rm d}E={\rm d}\Omega$. Notice
that the phase space volume $\Omega(E;\lambda)$ at fixed $\lambda$ corresponding to $E_{\text{min}}\equiv E_{\text{min}}(\lambda)$ is vanishing: Therefore we recast Eqn.~(\ref{eq:partA_01}) as

\begin{equation}
\int_{\Gamma}f\big(H({\bf p},{\bf q};\lambda)\big){\rm d}{\bf p}{\rm d}{\bf q}=\int_{0}^{\infty}f\big(E(\Omega;\lambda)\big){\rm d}\Omega\label{eq:equivalent}\;.
\end{equation}
This useful relation  is proven mathematically rigorously in Ref.~\cite{kasuga1961_1}.\\

\section{Bi-stochastic nature of $P(\bar{\Omega}_{\tau}|\bar{\Omega}_{0})$}
To determine the conditional probability for $\bar{\Omega}_{\tau}$ given the initial condition $\bar{\Omega}_{0}$,
i.e. $P(\bar{\Omega}_{\tau}|\bar{\Omega}_{0})$,
under the application of the protocol $\lambda_{0}\rightarrow\lambda_{\tau}$,
we consider first all the set of all phase space points $({\bf p}_{0},{\bf q}_{0})$ that
start with $\Omega(H({\bf p}_{0},{\bf q}_{0};\lambda_{0});\lambda_{0})=\bar{\Omega}_{0}$
and  analyze the final phase space volume $\Omega(H({\bf p}_{\tau},{\bf q}_{\tau};\lambda_{\tau});\lambda_{\tau})$,
where $({\bf p}_{\tau},{\bf q}_{\tau})$ denote the  time evolved Hamiltonian solutions starting at $({\bf p}_{0},{\bf q}_{0})$
with $\lambda_{0}\rightarrow\lambda_{\tau}$. The cumulative
probability of $P(\bar{\Omega}_{\tau}|\bar{\Omega}_{0})$ is
the conditional probability given $\bar{\Omega}_{0}$
at $\lambda_{0}$ to end up with $\Omega(H({\bf p}_{\tau},{\bf q}_{\tau};\lambda_{\tau});\lambda_{\tau})\le\bar{\Omega}_{\tau}$.
This cumulative conditional probability reads
\begin{widetext}
\begin{equation}
\int_{0}^{\bar{\Omega}_{\tau}}P(\Omega|\bar{\Omega}_{0}){\rm d}\Omega=\int_{\Gamma}\Theta\bigg(\bar{\Omega}_{\tau}-\Omega\big(H({\bf p}_{\tau},{\bf q}_{\tau};\lambda_{\tau});\lambda_{\tau}\big)\bigg)\frac{\delta\big(E(\bar{\Omega}_{0};\lambda_{0})-H({\bf p}_{0},{\bf q}_{0};\lambda_{0})\big)}{\omega\big(E(\bar{\Omega}_{0};\lambda_{0});\lambda_{0}\big)}{\rm d}{\bf p}_{0}{\rm d}{\bf q}_{0},\label{eq:cummulative}
\end{equation}
where the $\delta$-function stems from the normalized initial probability
on the phase space shell $\Omega = \bar{\Omega}_0$

\begin{equation}
\int_{\Gamma}\frac{\delta\big(E(\Omega;\lambda)-H({\bf p},{\bf q};\lambda)\big)}{\omega\big(E(\Omega;\lambda);\lambda\big)}{\rm d}{\bf p}{\rm d}{\bf q}=1.\label{eq:normalization}
\end{equation}
Upon taking the derivative of Eqn.~(\ref{eq:cummulative}), we obtain
\begin{equation}
P(\bar{\Omega}_{\tau}|\bar{\Omega}_{0})=\int_{\Gamma}\delta\bigg(\bar{\Omega}_{\tau}-\Omega\big(H({\bf p}_{\tau},{\bf q}_{\tau};\lambda_{\tau});\lambda_{\tau}\big)\bigg)\frac{\delta\big(E(\bar{\Omega}_{0};\lambda_{0})-H({\bf p}_{0},{\bf q}_{0};\lambda_{0})\big)}{\omega\big(E(\bar{\Omega}_{0};\lambda_{0});\lambda_{0}\big)}{\rm d}{\bf p}_{0}{\rm d}{\bf q}_{0},
\label{eq:conditional_probability}
\end{equation}
which  consistently obeys $P(\bar{\Omega}_{\tau}|\bar{\Omega}_{0})=\delta(\bar{\Omega}_{\tau}-\bar{\Omega}_{0})$, as $\tau\rightarrow 0^+$.
We readily observe the   normalization condition for $P(\bar{\Omega}_{\tau}|\bar{\Omega}_{0})$, namely

\begin{eqnarray}
\int_{0}^{\infty}P(\Omega|\bar{\Omega}_{0}){\rm d}\Omega & = & \lim_{\bar{\Omega}_{\tau}\rightarrow\infty}\int_{\Gamma}\Theta\bigg(\bar{\Omega}_{\tau}-\Omega\big(H({\bf p}_{\tau},{\bf q}_{\tau};\lambda_{\tau});\lambda_{\tau}\big)\bigg)\frac{\delta\big(E(\bar{\Omega}_{0};\lambda_{0})-H({\bf p}_{0},{\bf q}_{0};\lambda_{0})\big)}{\omega\big(E(\bar{\Omega}_{0};\lambda_{0});\lambda_{0}\big)}{\rm d}{\bf p}_{0}{\rm d}{\bf q}_{0}\nonumber\\
 & = & \int_{\Gamma}\frac{\delta\big(E(\bar{\Omega}_{0};\lambda_{0})-H({\bf p}_{0},{\bf q}_{0};\lambda_{0})\big)}{\omega\big(E(\bar{\Omega}_{0};\lambda_{0});\lambda_{0}\big)}{\rm d}{\bf p}_{0}{\rm d}{\bf q}_{0}=1.
\end{eqnarray}

We next demonstrate  the bi-stochastic property by integrating over $\bar{\Omega}_{0}$
in Eqn.~(\ref{eq:conditional_probability}); i.e.,
\begin{eqnarray}
\int_{0}^{\infty}P(\bar{\Omega}_{\tau}|\bar{\Omega}_{0}){\rm d}\bar{\Omega}_{0} & = & \int_{0}^{\infty}{\rm d}\bar{\Omega}_{0}\int_{\Gamma}\delta\bigg(\bar{\Omega}_{\tau}-\Omega\big(H({\bf p}_{\tau},{\bf q}_{\tau};\lambda_{\tau});\lambda_{\tau}\big)\bigg)\frac{\delta\big(E(\bar{\Omega}_{0};\lambda_{0})-H({\bf p}_{0},{\bf q}_{0};\lambda_{0})\big)}{\omega\big(E(\bar{\Omega}_{0};\lambda_{0});\lambda_{0}\big)}{\rm d}{\bf p}_{0}{\rm d}{\bf q}_{0} \nonumber \\
 & = & \int_{\Gamma}\delta\bigg(\bar{\Omega}_{\tau}-\Omega\big(H({\bf p}_{\tau},{\bf q}_{\tau};\lambda_{\tau});\lambda_{\tau}\big)\bigg){\rm d}{\bf p}_{0}{\rm d}{\bf q}_{0}\int_{0}^{\infty}\frac{\delta\big(E(\bar{\Omega}_{0};\lambda_{0})-H({\bf p}_{0},{\bf q}_{0};\lambda_{0})\big)}{\omega\big(E(\bar{\Omega}_{0};\lambda_{0});\lambda_{0}\big)}{\rm d}\bar{\Omega}_{0}.\label{eq:partB_01}
\end{eqnarray}
Observing that ${\rm d}\bar{\Omega}_{0}={\rm d}\Omega(\bar{E}_{0};\lambda_{0})=\omega(\bar{E}_{0};\lambda_{0}){\rm d}\bar{E}_{0}$,
and  writing $E(\bar{\Omega}_{0};\lambda_{0})=\bar{E}_{0}$, Eqn.~(\ref{eq:partB_01}) assumes the form
\begin{eqnarray}
\int_{0}^{\infty}P(\bar{\Omega}_{\tau}|\bar{\Omega}_{0}){\rm d}\bar{\Omega}_{0}
 & = & \int_{\Gamma}\delta\bigg(\bar{\Omega}_{\tau}-\Omega\big(H({\bf p}_{\tau},{\bf q}_{\tau};\lambda_{\tau});\lambda_{\tau}\big)\bigg){\rm d}{\bf p}_{0}{\rm d}{\bf q}_{0}\int_{0}^{\infty} \delta\big( \bar{E}_{0}-H({\bf p}_{0},{\bf q}_{0};\lambda_{0})\big){\rm d}\bar{E}_{0}\\
 & = & \int_{\Gamma}\delta\bigg(\bar{\Omega}_{\tau}-\Omega\big(H({\bf p}_{\tau},{\bf q}_{\tau};\lambda_{\tau});\lambda_{\tau}\big)\bigg){\rm d}{\bf p}_{0}{\rm d}{\bf q}_{0}.
\end{eqnarray}
The Jacobian of $({\bf p}_{0},{\bf q}_{0})\rightarrow({\bf p}_{\tau},{\bf q}_{\tau})$
equals unity, and making use of the relation in Eq. (\ref{eq:equivalent}) we find
\begin{eqnarray}
\int_{0}^{\infty}P(\bar{\Omega}_{\tau}|\bar{\Omega}_{0}){\rm d}\bar{\Omega}_{0} & = & \int_{\Gamma}\delta\bigg(\bar{\Omega}_{\tau}-\Omega\big(H({\bf p}_{\tau},{\bf q}_{\tau};\lambda_{\tau});\lambda_{\tau}\big)\bigg){\rm d}{\bf p}_{0}{\rm d}{\bf q}_{0}\\
 & = & \int_{\Gamma}\delta\bigg(\bar{\Omega}_{\tau}-\Omega\big(H({\bf p}_{\tau},{\bf q}_{\tau};\lambda_{\tau});\lambda_{\tau}\big)\bigg){\rm d}{\bf p}_{\tau}{\rm d}{\bf q}_{\tau}\\
 & = & \int_{0}^{\infty}\delta(\bar{\Omega}_{\tau}-\Omega){\rm d}\Omega=1,
\end{eqnarray}
\end{widetext}
Thus, this shows that the  bi-stochastic property
\begin{equation}
\int_{0}^{\infty}P(\bar{\Omega}_{\tau}|\bar{\Omega}_{0}){\rm d}\bar{\Omega}_{0}=1=\int_{0}^{\infty}P(\bar{\Omega}_{\tau}|\bar{\Omega}_{0}){\rm d}\bar{\Omega}_{\tau}\label{eq:bi-stochastic}
\end{equation}
indeed holds true for arbitrary work processes.


\section{Minimization of $\langle e^{-2\beta W}\rangle$ by Adiabatic Work Protocols }
As mentioned in the main text, the quantity $\langle e^{-2\beta W}\rangle$ can be written as
\begin{eqnarray}
\langle e^{-2\beta W}\rangle &=& \int_{0}^{\infty}{\rm d}\bar{\Omega}_{0}\int_{0}^{\infty}{\rm d}\bar{\Omega}_{\tau}e^{-2\beta[E(\bar{\Omega}_{\tau};\lambda_\tau)-E(\bar{\Omega}_{0};\lambda_{0})]} \nonumber \\
&&\times\ P(\bar{\Omega}_{\tau}|\bar{\Omega}_{0})\rho_{0}(\bar{\Omega}_{0}),\label{eq:rewrite}
\end{eqnarray}
where $\rho_{0}={e^{-\beta E(\bar{\Omega}_0;\lambda_0)}}/{Z(\beta; \lambda_0})$ denotes the initial canonical distribution in $\bar{\Omega}_0$-space.  Here, $P(\bar{\Omega}_{\tau}|\bar{\Omega}_{0})$ is the conditional probability to reach $\bar{\Omega}_{\tau}$, given initially $\bar{\Omega}_{0}$. Inspired by a  result obtained for ergodic one-dimensional systems obtained in  \cite{Gaoyang_Minimal_PhysRevE.92.022130}, we show  here
that the expression in Eq.~(\ref{eq:rewrite}) becomes minimized by any mechanical adiabatic process operating on an arbitrary chaotic Hamiltonian system.

Let us assume first that $\langle e^{-2\beta W}\rangle<\infty$. We now make use of an  integration by parts applied to Eq.~(\ref{eq:rewrite}), by rewriting $P(\bar{\Omega}_\tau|\bar{\Omega}_0)$ as a derivative. This  yields
\begin{widetext}
\begin{eqnarray}
\langle e^{-2\beta W}\rangle
& = & \int_{0}^{\infty} {\rm d}\bar{\Omega}_{0}e^{2\beta E(\bar{\Omega}_{0};\lambda_{0})}\rho_{0}(\bar{\Omega}_{0})\bigg(\int_{0}^{\infty}{\rm d}\bar{\Omega}_{\tau}e^{-2\beta E(\bar{\Omega}_{\tau};\lambda_{\tau})}\big[\frac{{\rm d}}{{\rm d}\bar{\Omega}_{\tau}}\int_{0}^{\bar{\Omega}_{\tau}}{\rm d}\Omega P(\Omega|\bar{\Omega}_{0})\big]\bigg)    \\
& = & \int_{0}^{\infty}{\rm d}\bar{\Omega}_{0}e^{2\beta E(\bar{\Omega}_{0};\lambda_{0})}\rho_{0}(\bar{\Omega}_{0})\bigg(\big[e^{-2\beta E(\bar{\Omega}_{\tau};\lambda_{\tau})}\int_{0}^{\bar{\Omega}_{\tau}}{\rm d}\Omega P(\Omega|\bar{\Omega}_{0})\big]\big|_{\bar{\Omega}_{\tau}=0}^{\infty}\nonumber \\
&&-\int_{0}^{\infty}{\rm d}\bar{\Omega}_{\tau} \big({{\rm d}e^{-2\beta E(\bar{\Omega}_{\tau};\lambda_{\tau})}}/{{\rm d}\bar{\Omega}_{\tau}}\big)\int_{0}^{\bar{\Omega}_{\tau}}{\rm d}\Omega P(\Omega|\bar{\Omega}_{0})\bigg).
\end{eqnarray}
Observing that the boundary terms at
at  $\bar{\Omega}_{\tau}=0$ or $\bar{\Omega}_{\tau}=\infty$ are vanishing, we end up with
\begin{equation}
\langle e^{-2\beta W}\rangle  =  -\int_{0}^{\infty}{\rm d}\bar{\Omega}_{\tau}\big({{\rm d}e^{-2\beta E(\bar{\Omega}_{\tau};\lambda_{\tau})}}/{{\rm d}\bar{\Omega}_{\tau}}\big)\bigg[\int_{0}^{\infty}{\rm d}\bar{\Omega}_{0}e^{2\beta E(\bar{\Omega}_{0};\lambda_{0})}\rho_{0}(\bar{\Omega}_{0})\int_{0}^{\bar{\Omega}_{\tau}}{\rm d}\Omega P(\Omega|\bar{\Omega}_{0})\bigg].
\label{eq1}
\end{equation}

Next, let $\mathcal{A}(\bar{\Omega}_{\tau})$
denote the integral part inside the big square brackets in Eq.~(\ref{eq1}). Because $E(\Omega;\lambda)$ monotonically increases
with $\Omega$, one has $e^{\beta E(\bar{\Omega}_0;\lambda_0)}> e^{\beta E(\bar{\Omega}_\tau;\lambda_0)}$ if $\bar{\Omega}_0>\bar{\Omega}_\tau$. Exploiting this fact, we can write
\begin{eqnarray}
\mathcal{A}(\bar{\Omega}_{\tau}) & = &  \int_{0}^{\bar{\Omega}_{\tau}}{\rm d}\Omega\int_{0}^{\bar{\Omega}_{\tau}}{\rm d}\bar{\Omega}_{0}\frac{e^{\beta E(\bar{\Omega}_{0};\lambda_{0})}}{Z(\beta;\lambda_{0})}P(\Omega|\bar{\Omega}_{0})+\int_{0}^{\bar{\Omega}_{\tau}}{\rm d}\Omega\int_{\bar{\Omega}_{\tau}}^{\infty}{\rm d}\bar{\Omega}_{0}\frac{e^{\beta E(\bar{\Omega}_{0};\lambda_{0})}}{Z(\beta;\lambda_{0})}P(\Omega|\bar{\Omega}_{0}) \\
 & \ge & \int_{0}^{\bar{\Omega}_{\tau}}{\rm d}\Omega\int_{0}^{\bar{\Omega}_{\tau}}{\rm d}\bar{\Omega}_{0}\frac{e^{\beta E(\bar{\Omega}_{0};\lambda_{0})}}{Z(\beta;\lambda_{0})}P(\Omega|\bar{\Omega}_{0})+\frac{e^{\beta E(\bar{\Omega}_{\tau};\lambda_{0})}}{Z(\beta;\lambda_{0})}\int_{0}^{\bar{\Omega}_{\tau}}{\rm d}\Omega\int_{\bar{\Omega}_{\tau}}^{\infty}{\rm d}\bar{\Omega}_{0}P(\Omega|\bar{\Omega}_{0})\\
 & = & \int_{0}^{\bar{\Omega}_{\tau}}{\rm d}\Omega\int_{0}^{\bar{\Omega}_{\tau}}{\rm d}\bar{\Omega}_{0}\frac{e^{\beta E(\bar{\Omega}_{0};\lambda_{0})}}{Z(\beta;\lambda_{0})}P(\Omega|\bar{\Omega}_{0}) \nonumber\\
 &&+\ \frac{e^{\beta E(\bar{\Omega}_{\tau};\lambda_{0})}}{Z(\beta;\lambda_{0})}\bigg[\int_{0}^{\bar{\Omega}_{\tau}}{\rm d}\Omega\int_0^{\infty}{\rm d}\bar{\Omega}_{0}P(\Omega|\bar{\Omega}_{0})-\int_{0}^{\bar{\Omega}_{\tau}}{\rm d}\Omega\int_{0}^{\bar{\Omega}_{\tau}}{\rm d}\bar{\Omega}_{0}P(\Omega|\bar{\Omega}_{0})\bigg].
\end{eqnarray}
Using Eq.~(\ref{eq:bi-stochastic}), this equation can be reformed further by writing
\begin{eqnarray}
\mathcal{A}(\bar{\Omega}_{\tau})  & \ge &  \int_{0}^{\bar{\Omega}_{\tau}}{\rm d}\Omega\int_{0}^{\bar{\Omega}_{\tau}}{\rm d}\bar{\Omega}_{0}\frac{e^{\beta E(\bar{\Omega}_{0};\lambda_{0})}}{Z(\beta;\lambda_{0})}P(\Omega|\bar{\Omega}_{0}) + \frac{e^{\beta E(\bar{\Omega}_{\tau};\lambda_{0})}}{Z(\beta;\lambda_{0})}\bigg[\bar{\Omega}_{\tau} - \int_{0}^{\bar{\Omega}_{\tau}}{\rm d}\Omega\int_{0}^{\bar{\Omega}_{\tau}}{\rm d}\bar{\Omega}_{0}P(\Omega|\bar{\Omega}_{0})\bigg]\\
 & = & \int_{0}^{\bar{\Omega}_{\tau}}{\rm d}\Omega\int_{0}^{\bar{\Omega}_{\tau}}{\rm d}\bar{\Omega}_{0}\frac{e^{\beta E(\bar{\Omega}_{0};\lambda_{0})}}{Z(\beta;\lambda_{0})}P(\Omega|\bar{\Omega}_{0}) \nonumber\\
 &  & +\ \frac{e^{\beta E(\bar{\Omega}_{\tau};\lambda_{0})}}{Z(\beta;\lambda_{0})}\bigg[\int_{0}^{\infty}{\rm d}\Omega\int_{0}^{\bar{\Omega}_{\tau}}{\rm d}\bar{\Omega}_{0}P(\Omega|\bar{\Omega}_{0})-\int_{0}^{\bar{\Omega}_{\tau}}{\rm d}\Omega\int_{0}^{\bar{\Omega}_{\tau}}{\rm d}\bar{\Omega}_{0}P(\Omega|\bar{\Omega}_{0})\bigg] \\
 & \ge & \int_{0}^{\bar{\Omega}_{\tau}}{\rm d}\Omega\int_{0}^{\bar{\Omega}_{\tau}}{\rm d}\bar{\Omega}_{0}\frac{e^{\beta E(\bar{\Omega}_{0};\lambda_{0})}}{Z(\beta;\lambda_{0})}P(\Omega|\bar{\Omega}_{0})+\int_{\bar{\Omega}_{\tau}}^{\infty}{\rm d}\Omega\int_{0}^{\bar{\Omega}_{\tau}}{\rm d}\bar{\Omega}_{0}\frac{e^{\beta E(\bar{\Omega}_{0};\lambda_{0})}}{Z(\beta;\lambda_{0})}P(\Omega|\bar{\Omega}_{0})\\
 & = & \int_{0}^{\bar{\Omega}_{\tau}}{\rm d}\bar{\Omega}_{0}\frac{e^{\beta E(\bar{\Omega}_{0};\lambda_{0})}}{Z(\beta;\lambda_{0})}\int_{0}^{\infty}{\rm d}\Omega P(\Omega|\bar{\Omega}_{0})\\
 & = & \int_{0}^{\bar{\Omega}_{\tau}}{\rm d}\bar{\Omega}_{0}\frac{e^{\beta E(\bar{\Omega}_{0};\lambda_{0})}}{Z(\beta;\lambda_{0})}.
\end{eqnarray}
Inserting this  partial finding  into Eqn.~(\ref{eq1}) and using once again an integration by parts, we obtain
\begin{eqnarray}
\langle e^{-2\beta W}\rangle & = & -\int_{0}^{\infty}{\rm d}\bar{\Omega}_{\tau}\big({{\rm d}e^{-2\beta E(\bar{\Omega}_{\tau};\lambda_{\tau})}}/{{\rm d}\bar{\Omega}_{\tau}}\big)\mathcal{A}(\bar{\Omega}_{\tau})  \\
&\ge&   -\int_{0}^{\infty}{\rm d}\bar{\Omega}_{\tau}\big({{\rm d}e^{-2\beta E(\bar{\Omega}_{\tau};\lambda_{\tau})}}/{{\rm d}\bar{\Omega}_{\tau}}\big)\int_{0}^{\bar{\Omega}_{\tau}}{\rm d}\bar{\Omega}_{0}\frac{e^{\beta E(\bar{\Omega}_{0};\lambda_{0})}}{Z(\beta;\lambda_{0})}\\
 & = & -\bigg(\big[e^{-2\beta E(\bar{\Omega}_{\tau};\lambda_{\tau})}\int_{0}^{\bar{\Omega}_{\tau}}{\rm d}\bar{\Omega}_{0}\frac{e^{\beta E(\bar{\Omega}_{0};\lambda_{0})}}{Z(\beta;\lambda_{0})}\big]\big|_{\bar{\Omega}_{\tau}=0}^{\infty}-\int_{0}^{\infty}{\rm d}\bar{\Omega}_{\tau}e^{-2\beta E(\bar{\Omega}_{\tau};\lambda_{\tau})}\big[\frac{{\rm d}}{{\rm d}\bar{\Omega}_{\tau}}\int_{0}^{\bar{\Omega}_{\tau}}{\rm d}\bar{\Omega}_{0}\frac{e^{\beta E(\bar{\Omega}_{0};\lambda_{0})}}{Z(\beta;\lambda_{0})}\big]\bigg)\\
 & = & \int_{0}^{\infty}{\rm d}\bar{\Omega}_{\tau}e^{-2\beta E(\bar{\Omega}_{\tau};\lambda_{\tau})}\frac{e^{\beta E(\bar{\Omega}_{\tau};\lambda_{0})}}{Z(\beta;\lambda_{0})}\\
 & = & \int_{0}^{\infty}{\rm d}\bar{\Omega}_{\tau}\int_{0}^{\infty}{\rm d}\bar{\Omega}_{0}e^{-2\beta[E(\bar{\Omega}_{\tau};\lambda_{\tau})-E(\bar{\Omega}_{0};\lambda_{0})]}\rho_{0}(\bar{\Omega}_{0})\delta(\bar{\Omega}_{\tau}-\bar{\Omega}_{0})\\
 & \equiv & \langle e^{-2\beta W_{{\rm ad}}}\rangle \;.
\end{eqnarray}
\end{widetext}
This shows that among all work protocols the minimal value of $\langle e^{-2\beta W}\rangle$ is assumed for
 $P(\bar{\Omega}_\tau|\bar{\Omega}_0)$ becoming the delta function $\delta(\bar{\Omega}_{\tau}-\bar{\Omega}_{0}) $.  This precisely refers to a situation arising from using
mechanically  adiabatic work protocols in chaotic systems; only then we have that the associated total phase space volume $\Omega(E;\lambda)$ is an  adiabatic invariant.

\section{Behavior of ${\rm Var}(e^{-\beta W})$ in a non-ergodic system}
 In order to see if ${\rm Var}(e^{-\beta W})$ may increase as the duration of a work protocol increases in non-ergodic systems,
 we have also considered a two-dimensional nonlinear oscillator system in dimensionless units.
The Hamiltonian is given by
\begin{equation}
H(t)=\frac{p_{x}^{2}}{2m}+\frac{p_{y}^{2}}{2m}+\frac{1}{2}m\omega_{x}^{2}(t)x^{2}+\frac{1}{2}m\omega_{y}^{2}y^{2}+\lambda x^{2}y^{2},
\label{Heq}
\end{equation}
with $m=1$, $\omega_{y}=1.07$ and $\lambda=0.05$. The work protocol is given by $
\omega_{x}=(\omega_{\tau}-\omega_{0})\sin(\pi\frac{t}{\tau})+\omega_{0}$,
with $\omega_{0}=1.0$, $\omega_{\tau}=1.1$. Simulation is done based on $150,000$ trajectories for each $\tau$, with $\beta=0.1$.
Results are presented in Fig.~\ref{fig:add} below.  It is seen from Fig.~~\ref{fig:add} that, for a non-ergodic system, ${\rm Var}( e^{-\beta W})$ associated with a fast work protocol may be much smaller than
that for a very slow protocol. In particular,
for $\tau=0.01$, ${\rm Var}(e^{-\beta W})\sim 1.4 \times 10^{-6}$; For $\tau=1000$, ${\rm Var}(e^{-\beta W})\sim8.9\times 10^{-4}$, namely, the variance in the exponential work is a few hundred times larger as $\tau$ increases.  This observation is in contrast to our general theoretical prediction exclusively for ergodic systems. 
\begin{figure} 
\centering      
\includegraphics[width=0.7\linewidth]{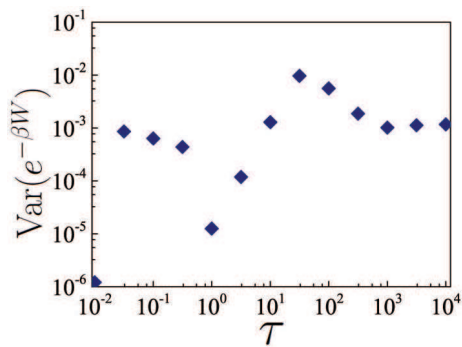}
\caption{(color online) Variance of exponential work, denoted ${\rm Var}(e^{-\beta W})$,
 vs the duration $\tau$ of the work protocol mentioned under Eq.~(\ref{Heq}), for a classical system of mixed phase space structure. Note the logarithmic scale used for both $\tau$ and ${\rm Var}(e^{-\beta W})$.}
\label{fig:add}
\end{figure}


\section{Implications of a diverging second moment  $\langle e^{-2\beta W_{\text{ad}}}\rangle$}
A generalization of the central limit theorem (CLT) \cite{durrett, kolmogorov}  states the following:
Suppose $X_{1}$, $X_{2}$, ... are independent identical distributed (i.i.d.) random variables.
Let $X$  represent any such random variable which is assumed to obey the two conditions:

\begin{enumerate}
\item $\lim_{x\to\infty}P(X>x)/P(|X|>x)=\gamma \in[0,1]$
\item $P(|X|>x)=x^{-\alpha}L(x)$, where $0<\alpha<2$ and $L(x)$ being slowly varying; i.e., $\lim_{x\to\infty}L(tx)/L(x)=1$,
for all $t>0$.
\end{enumerate}
Here, $P(A)$ denotes the probability of the event $A$.
It then follows that as $n\to\infty$,
\begin{equation}
\frac{S_{n}-b_{n}} {a_{n}}\Rightarrow Y,\label{eq:weak_converge_to_Y}
\end{equation}
where $Y$ possesses a non-degenerate distribution, wherein
\begin{eqnarray}
S_{n}&=&\sum_{i=1}^{n}X_{i},\label{eq:def_S_N}\\
a_{n}&=&\inf\{x\vcentcolon P(|X|>x)\le n^{-1}\},\label{eq:def_a_N}\\
b_{n}&=&n\langle X\Theta(a_{n}-|X|)\rangle.\label{eq:def_b_N}
\end{eqnarray}
with $\langle\cdot\rangle$ denoting the statistical expectation value,
and $\Theta$ denoting the unit step function.
That is, $\langle X \Theta(a_{n}-|X|)\rangle$ represents the expectation value
of $X$ truncated at $\pm a_{n}$.

It is worth noting that the above generalized CLT-theorem should reduce to the commonly known CLT if $\alpha>2$, with $\lim_{n\rightarrow \infty}a_{n}\rightarrow n^{1/2}$ and $\lim_{n\rightarrow \infty}b_{n}\rightarrow n\langle X\rangle$, while the random variable $Y$ becomes a Gaussian distributed random variable with a vanishing average  and sharing the same finite variance ${\rm Var}(Y) = {\rm Var}(X_i)$ as $X_i$.

To connect the above theorem with exponential work fluctuations, we  set $X_i=e^{-\beta W_i}$. Here $W_i$ is the random work from an $i$-th measurement with initial condition randomly picked from the canonically distributed phase space; i.e.,  $W_i$ are i.i.d. random variables just as the set $X_i$ above. It readily follows  that $\lim_{x\to\infty}P(X >x)/P(|X |>x)=\gamma \in[0,1]$
with $\gamma=1$, because $X =e^{-\beta W}$ is positive definite.  For the Sinai billiard systems considered in the main text undergoing an adiabatic protocol,
$W \equiv W_{\rm ad}=\bar{E}_0({\lambda_0}-{\lambda_\tau})/{\lambda_\tau}=\bar{\Omega}_0({\lambda_0}-{\lambda_\tau})/(2\pi m \lambda_0{\lambda_\tau})$ (see the main text for the notation), where $\bar{E}_0$ and $\bar{\Omega}_0$ denote  the energy and phase space volume of the initial state, respectively. We use here that $\bar{\Omega}_0=2\pi m \lambda_0\bar{E}_0$ for the Sinai billiard model. In the regime under our consideration here we have that  $\lambda_\tau > 2 \lambda_0$; thus  $W_{\rm ad}$ is always negative, because $\lambda > 0$ denotes the free area, being strictly positive.
Let $x= e^{-\beta(\frac{\lambda_0}{\lambda_\tau}-1){r}/({2\pi m \lambda_0})} $.  Then the probability  $P(|X|>x)$
is  given by the probability of finding $\bar{\Omega}_0 >r $; i.e., it is given by the tail of the initial canonical probability distribution.   Explicitly,
\begin{eqnarray}
P(|X|>x) &=&\int_{|X|>x}  \frac{e^{-\beta H({\bf p}_0,{\bf q}_0;\lambda_0)}}{Z(\beta;\lambda_0)} {\rm d}{\bf p}_0 {\rm d}{\bf q}_0 \\
&=&  \int_{r}^{+\infty}  \frac{\beta}{2\pi m \lambda_0}e^{-\beta {{\bar{\Omega}}_0}/{(2\pi m \lambda_0)}} {\rm d} \bar{\Omega}_0 \\
& =&  e^{-\beta r / (2\pi m \lambda_0)}   \\
 & = & x^{-{\lambda_\tau}/({\lambda_\tau-\lambda_0})} \\
 & = & x^{-{s}/{s-1}} \;.
\end{eqnarray}
Here, $s = \lambda_\tau/\lambda_0$  and we made use of the relation in  Eq.~(\ref{eq:equivalent}) and, as well, that $Z(\beta;\lambda_0)=2\pi m \lambda_0/\beta$ for those two-dimensional Sinai billiards.

With  $\lambda_{\tau} > 2\lambda_0$ (i.e. $s>2$), yielding  $\alpha = {s}/{(s-1)}<2$, the above-mentioned condition in (2.) for the generalized CLT is  satisfied with
$L(x)\equiv 1$. Referring to Eq.~(\ref{eq:def_a_N}) we obtain
\begin{eqnarray}
a_{n} & = & \inf\{x\vcentcolon |x|^{-\alpha}\le n^{-1}\}  \\
 & = & n^{1/\alpha}\label{eq:a_n}.
\end{eqnarray}

Moreover, $b_{n}\rightarrow n\langle X\rangle$ as $n \rightarrow \infty$.  With these intermediate findings , Eq.~(\ref{eq:weak_converge_to_Y}) becomes
\begin{equation}
\label{eqa}
\lim_{n\to\infty}\frac{\big(\sum_{i=1}^{n}e^{-\beta W_i}\big)-n\langle X\rangle} {n^{1/\alpha}}\Rightarrow Y \;.
\end{equation}
Upon dividing  both sides of Eq.~(\ref{eqa}) by ${n^{1-\frac{1}{\alpha}}}$ and with $\alpha = {s}/{(s-1)}$ we arrive at
\begin{equation}
\lim_{n\to\infty}\bigg(\sum_{i=1}^{n}\frac{e^{-\beta W_{i}}}{n}\bigg)-\langle e^{-\beta W_{\rm ad}}\rangle \Rightarrow\frac{Y}{n^{{1}/{s}}}\;,
\label{eq:weak_convergence}
\end{equation}
wherein $Y$ denotes the  limiting ($n \rightarrow \infty $) random variable for  exponential adiabatic work.
Equation~(\ref{eq:weak_convergence})  describes how the error between the statistical estimate for the average of adiabatic random work values, i.e.,
$\sum_{i=1}^{n}{e^{-\beta W_{i}}}/{n}$, and the measure-theoretic average itself, i.e.,  $\langle e^{-\beta W_{\rm ad}}\rangle$, scales with $n$ in the limit $n \rightarrow \infty$. Put differently, the coefficient $n^{{1}/{s}}$ on the right hand side of Eq.~(\ref{eq:weak_convergence}) indicates how the error {\it scales} with increasing  $n$.   Note that this error scaling law $n^{-{1}/{s}}$ depends strongly on $s={\lambda_{\tau}}/{\lambda_{0}}$.  This latter dependence describes an intriguing  protocol-dependent feature.




\begin{thebibliography}{100}
\bibitem{note2}Other protocols such as a linear protocol $\lambda(t)=\lambda_0+(\lambda_\tau-\lambda_0)t/\tau$  are also checked and all our results (not shown) are essentially the same.
\bibitem{footnote}A careful analysis shows that ${\rm Var}(e^{-\beta W})$ simulated for the integrable
case (namely, a particle moving in a rectangular square billard) is not converged yet because the second moment of the variance, determined by $\langle e^{- 4\beta W}\rangle $, diverges
\end{thebibliography}
\end{document}